\documentclass[12pt,twoside,a4paper]{article}
\usepackage{epsfig}
\usepackage{amssymb}
\author{Martin Reuter and Jan-Markus Schwindt
 \footnote{ E-mail: reuter@thep.physik.uni-mainz.de, 
 schwindt@thep.physik.uni-mainz.de}}
\date{} 

\title{A Minimal Length from the Cutoff Modes in Asymptotically Safe 
Quantum Gravity}

\begin{document}
\maketitle 

\vspace{-8cm}
\begin{flushright} 
MZ-TH/05-23
\end{flushright}
\vspace{6cm}
\centerline{\small\it Institute of Physics, University of Mainz,}
\centerline{\small\it Staudingerweg 7, D-55128 Mainz, Germany}

\vspace{0.7cm} 

\begin{abstract}
 Within asymptotically safe Quantum Einstein Gravity (QEG), the quantum 4-sphere
 is discussed as a specific example of a fractal spacetime manifold. The
 relation between the infrared cutoff built into the effective average action
 and the corresponding coarse graining scale is investigated. Analyzing the 
 properties of the pertinent cutoff modes, the possibility that QEG generates
 a minimal length scale dynamically is explored. While there exists no minimal 
 proper length, the QEG sphere appears to be ``fuzzy" in the sense that there is
 a minimal angular separation below which two points cannot be resolved by the 
 cutoff modes.
\end{abstract}

\newpage
\noindent{\bf\large 1. Introduction}

It is an old speculation, based upon various heuristic arguments \cite{garay},
that quantum gravity induces a lower bound on physically realized distances.
Since this issue can be addressed only in a fundamental quantum theory of gravity
(as opposed to a low energy effective theory) it is natural to analyze it within 
Quantum Einstein Gravity (QEG). This theory is an attempt at the nonperturbative 
construction of a predictive quantum field theory of the metric tensor by means 
of a non-Gaussian renormalization group (RG) fixed point \cite{wein}-\cite{max}.
From what is known today it appears indeed increasingly likely that there 
does exist
an appropriate fixed point which renders QEG nonperturbatively renormalizable
or ``asymptotically safe" \cite{souma,oliver1,oliver2,oliver3}.

An important tool in analyzing the RG flow of QEG is the effective average action
and its exact functional RG equation \cite{avact,ym}. In the case of QEG \cite{mr},
the average action is a diffeomorphism invariant functional of the metric,
$\Gamma _k [g_{\mu\nu}]$, which depends on a variable infrared (IR) cutoff $k$.
For $k \rightarrow\infty $ it approaches the bare action $S$, while it equals the 
standard effective action at $k=0$. The effective field equations implied by 
$\Gamma _k$ define a $k$-dependent expectation value of the metric, a kind of mean
field, $\left < g_{\mu\nu}\right >_k$:
\begin{equation}\label{fe}
 \frac{\delta\Gamma _k}{\delta g_{\mu\nu}(x)}\left [ \left < g 
 \right >_k \right ]=0.
\end{equation} 

At least when one applies the average action technique to Euclidean non-gauge 
theories on flat space, i.e. when the metric is non-dynamical, or to statistical 
mechanical systems on a rigid lattice, the action $\Gamma _k$ has the following
properties \cite{avactrev}: (i) It defines an effective field theory at the 
momentum scale $k$. This means that every physical process which involves only a 
single momentum scale, say $p$, is well described by a tree level evaluation of
$\Gamma _k$ with $k$ chosen as $k=p$. (ii) At least heuristically \cite{avactrev},
$\Gamma _k$ may be interpreted as arising from a continuum version of a
Kadanoff-Wilson block spin procedure, i.e. it defines the dynamics of ``coarse
grained" dynamical variables which are averaged over a certain region of
Euclidean spacetime. Denoting the typical linear extension of the averaging region 
by $\ell$, one has $\ell \approx \pi /k$ in flat spacetime (at least for non-gauge
theories). In this sense, $\Gamma _k$ can be thought of as a ``microscope" with an
adjustable resolving power $\ell =\ell(k)$.

In quantum gravity where the metric is dynamical 
it is not clear a priori to what extent (i) and (ii) continue to be valid. 
For sure the relationship between the IR
cutoff $k$ and the ``averaging scale" $\ell$ is more complicated in general. It
is one of the purposes of the present paper to give a concrete definition of 
an ``averaging" or ``coarse graining" scale $\ell$, and to understand
how it depends on $k$. Using this definition, along with certain qualitative 
properties of the RG trajectories of QEG, we shall demonstrate that the theory 
generates a minimal length scale in a dynamical way. The interpretation of this
scale is rather subtle, however. One has to carefully distinguish different 
physical questions one could ask, because depending on the question a minimal length
will, or will not become visible.

Let us assume we have solved the exact RG equation and picked a specific RG 
trajectory, a curve $k \mapsto \Gamma _k$ on theory space. Then we can 
write down the effective Einstein equations (\ref{fe}) along this trajectory and,
after fixing appropriate symmetry and boundary conditions, 
find the corresponding family of mean field metrics
$\left \{ \left < g_{\mu\nu}\right > _k ;\; 0 \leq k< \infty\right \}$.
As for the interpretation of the average action approach in QEG, it is crucial to 
appreciate the fact that the infinitely many equations in (\ref{fe}), one at
each scale $k$, are valid simultaneously, and that all the mean fields 
$\left < g_{\mu\nu} \right >_k$ refer to one and the same physical system,
a ``quantum spacetime" in the QEG sense. The mean fields $\left < g_{\mu\nu} \right >_k$
describe the metric structure in dependence on the length scale on which the
spacetime manifold is probed. An observer exploring the structure of spacetime
using a ``microscope" of resolution $\ell(k)$ will perceive the universe as a
Riemannian manifold with the metric $\left < g_{\mu\nu} \right >_k$. While
$\left < g_{\mu\nu} \right >_k$ is a smooth classical metric at every fixed $k$,
the quantum spacetime can have fractal properties because on different scales
different metrics apply. In this sense the metric structure on the quantum
spacetime is given by an infinite set $\left \{ \left < g_{\mu\nu} \right >_k
;\; 0 \leq k< \infty\right \}$ of ordinary metrics. 

Recently it has been shown \cite{oliverfrac} that in asymptotically safe theories
of gravity, at sub-Planckian distances, spacetime is indeed a fractal whose spectral
dimension \cite{avra} equals 2. It is quite remarkable that a similar dynamical 
dimensional reduction from 4 macroscopic to 2 microscopic dimensions has also been
observed in Monte Carlo simulations of causal dynamical triangulations
\cite{ajl1,ajl2,ajl34}.
\footnote{See also \cite{nino} for related discussions of fractal spacetimes
within different theories of gravity.}

In order to understand the relation between $\ell$ and the IR cutoff $k$ we must
recall the essential steps in the construction of the average action \cite{mr}.
The formal starting point is the path integral $\int {\cal D}\gamma _{\mu\nu}
\exp \left ( -S[\gamma]\right )$ over all metrics $\gamma _{\mu\nu}$, gauge fixed by
means of a background gauge fixing condition \cite{back}. Even without an IR cutoff,
upon introducing sources and performing the usual Legendre 
transform one is led to an effective action $\Gamma\left [ g_{\mu\nu};
\bar{g}_{\mu\nu}\right ]$ which depends on two metrics, the expectation value of 
$\gamma _{\mu\nu}$, denoted $g_{\mu\nu}$, and the non-dynamical background field
$\bar{g}_{\mu\nu}$. It is well-known \cite{back} that the functional 
$\Gamma[g_{\mu\nu}]\equiv\Gamma[g_{\mu\nu};\bar{g}_{\mu\nu}=g_{\mu\nu}]$ 
obtained by equating the two
metrics generates the 1PI Green's functions of the theory.

The IR cutoff is implemented by first expanding the shifted integration variable
$h_{\mu\nu}\equiv\gamma _{\mu\nu}-\bar{g}_{\mu\nu}$ 
in terms of eigenmodes of $\bar{D}^2$,
the covariant Laplacian formed with the background metric 
$\bar{g}_{\mu\nu}$, and interpreting ${\cal D}h_{\mu\nu}$ as an
integration over all expansion coefficients.
Then a suppression term is introduced which damps the
contribution of all $\bar{D}^2$-modes with eigenvalues smaller than $k^2$.
Following the usual steps \cite{avactrev,bagber} this leads to the scale dependent
functional $\Gamma_k[g_{\mu\nu};\bar{g}_{\mu\nu}]$, and again the action with one
argument is obtained by equating the two metrics:
$\Gamma_k[g_{\mu\nu}]\equiv\Gamma_k[g_{\mu\nu};\bar{g}_{\mu\nu}=g_{\mu\nu}]$.
It is this action which appears in (\ref{fe}). Because of the identification
of the two metrics it is, in a sense, the eigenmodes of
$D^2$, constructed from the argument of $\Gamma_k[g]$, which are cut off at $k^2$.
Note that neither the $g_{\mu\nu}$- nor the $\bar{g}_{\mu\nu}$-argument of $\Gamma _k$
has any dependence on $k$. Therefore $\gamma _{\mu\nu}$ is expanded in terms of
eigenfunctions of a {\it fixed} operator $\bar{D}^2$. Since its eigenfunctions are
complete, we really integrate over all metrics when we lower $k$ from infinity
to zero.

Note that a $k$-dependent mean field arises only 
at the point where we go ``on shell" with $g_{\mu\nu}
=\bar{g}_{\mu\nu}$: the solution $\left < g_{\mu\nu} \right >_k$ to eq.~(\ref{fe})
depends on $k$, simply because $\Gamma _k$ does so.

In ref.~\cite{oliverfrac} an algorithm was proposed which allows the reconstruction
of the ``averaging" scale $\ell$ from the cutoff $k$. The input data is the set of 
metrics characterizing a quantum manifold, $\left \{ \left < g_{\mu\nu}\right >_k
\right \}$. The idea is to deduce the relation $\ell = \ell(k)$ from the spectral
properties of the {\it scale dependent} Laplacian ${\bf \Delta} _k \equiv D^2 \left ( 
\left < g_{\mu\nu}\right >_k \right )$ built with the solution of the effective
field equation. More precisely, for every fixed value of $k$, one solves the
eigenvalue problem of $-{\bf \Delta} _k$ and studies in particular the properties of the
eigenfunctions whose eigenvalue is $k^2$, or nearest to $k^2$ in the case of a discrete
spectrum. We shall refer to an
eigenmode of $-{\bf \Delta} _k$ whose eigenvalue is (approximately)
the square of the cutoff $k$ as a ``cutoff mode" (COM) 
and denote the set of all COMs by {\sf COM}($k$).

If we ignore the $k$-dependence of ${\bf \Delta} _k$ for a moment (as it would be
appropriate for matter theories in flat space) the COMs are, for a sharp cutoff,
precisely the last modes integrated out when lowering the cutoff, since the 
suppression term in the path integral cuts out all $h_{\mu\nu}$-modes with
eigenvalue smaller than $k^2$.

For a non-gauge theory in flat space the coarse graining or averaging of fields 
is a well defined procedure, based upon ordinary Fourier analysis,
and one finds that in this case the length $\ell$
is essentially the wave length of the last modes integrated out, the COMs.

This observation motivates the following 
tentative {\it definition} of $\ell$ in quantum gravity.
We determine the COMs of $-{\bf \Delta} _k$, analyze how fast these eigenfunctions vary
on spacetime, and read off a typical coordinate distance $\Delta x^\mu$
characterizing the scale on which they vary. For an oscillatory COM, for example,
$\Delta x$ would correspond to an oscillation period. (In general there is a certain 
freedom in the precise identification of the $\Delta x^\mu$ belonging to a 
specific cutoff mode. This ambiguity can be resolved by refining the definition
of $\Delta x^\mu$ on a case-by-case basis only. Since in the present paper only
oscillatory eigenfunctions will be encountered we shall not be more specific here.)
Finally we use the metric
$\left < g_{\mu\nu} \right >_k$ itself in order to convert $\Delta x^\mu$ to a
proper length. This proper length, by definition, is $\ell$.
Repeating the above steps for all values of $k$, we end up with a function
$\ell =\ell(k)$.
In general one will find that $\ell$ depends on the position on the manifold 
as well as on the direction of $\Delta x^\mu$. Since this position and direction
dependence plays no role in the present paper we shall not indicate it notationally.

In the following $\ell$ will always denote the intrinsic length scale of the
COMs obtained from the above algorithm. Our experience with theories in flat spacetime
suggests that the COM scale $\ell$ is a plausible {\it candidate} for a physically 
sensible resolution function $\ell =\ell(k)$, but there might also be others,
depending on the experimental setup one has in mind.
Interestingly enough, we shall see that in quantum gravity the COM scale $\ell(k)$
has properties which are quite different from classical physics.

For later comparison with QEG it is instructive to go through the various steps of the 
algorithm in the standard case.
We consider a classical flat Euclidean space with metric $\delta _{ij}$ and Laplacian
${\bf \Delta} =\sum \partial _i^2$. 
The eigenmodes of $-{\bf \Delta}$ with eigenvalue $k^2$ are
$\sin (\vec{q}\cdot\vec{x})$ and $\cos(\vec{q}\cdot\vec{x})$ with $\vec{q}^2=k^2$.
They vary between their maximal value $+1$ and their minimal value $-1$ in a 
coordinate interval $\Delta\vec{x}$ with $(\Delta\vec{x})^2=\pi^2/k^2$.
The length corresponding to this coordinate distance, obtained with the metric
$\delta _{ij}$, is
\begin{equation}
\ell(k)=\sqrt{\delta _{ij}\Delta x^i \Delta x^j}=\pi /k.
\end{equation}
Thus we recover the expected inverse proportionality $\ell\propto 1/k$.

The most difficult step in the construction of QEG spacetimes consists in finding the
RG trajectories. The running action $\Gamma_k[g_{\mu\nu}]$ satisfies an
exact functional RG equation \cite{mr}. In practice it is usually solved on a
truncated theory space. In the Einstein-Hilbert truncation, for instance, 
$\Gamma_k$ is approximated by a functional of the form
\begin{eqnarray}
\label{3in2}
\Gamma_k[g]=\left(16\pi G(k)\right)^{-1}\int d^4x\,\sqrt{g}\left\{
-R(g)+2 \Lambda (k)\right\}
\end{eqnarray}
involving a running Newton constant $G(k)$ and cosmological constant 
$\Lambda (k)$.

For each $k$, the action (\ref{3in2}) implies an effective field equation
\begin{equation}\label{fieldeq}
 R_{\mu\nu}(\left < g \right > _k)=\Lambda (k) \left < g_{\mu\nu}\right > _k.
\end{equation}  
Note that the running Newton constant $G(k)$ does not appear in this effective
Einstein equation. It enters only when matter fields are introduced.

\begin{figure}
\centerline{\psfig{figure=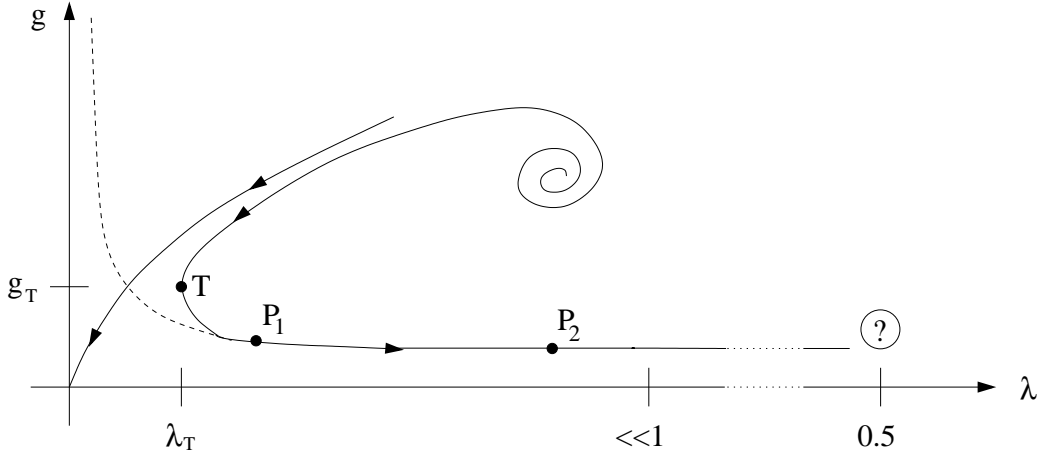, width=5.4in}}
\caption{A Type IIIa trajectory and the separatrix. The dashed line is a
trajectory of the canonical RG flow. The arrows point in the direction of 
decreasing $k$.}
\label{traj}
\end{figure}

The qualitative properties of the trajectories following from the Einstein-Hilbert
approximation are well-known by now \cite{frank1}.
Fig.~1 shows a ``Type IIIa" trajectory which would be the type that is 
presumably realized in the real universe since it is the only type that has a 
positive Newton's constant $G(k)$ and a small positive cosmological constant 
$\Lambda(k)$ at macroscopic scales. In Fig.~1 it is plotted in terms of the
dimensionless parameters 
$g(k) \equiv k^2 G(k)$ and $\lambda (k) \equiv \Lambda (k) /k^2$ 
and compared to the canonical
trajectory of classical gravity (dashed curve) with $\Lambda =$const and $G=$const.
The Type IIIa trajectory contains the following four parts, with
increasing values of the cutoff $k$:  \\ \\
i) The classical regime for small $k$ where the trajectory 
is identical to the canonical one.
(In Fig.~1 the segment between the points $P_1$ and $P_2$.)\\
ii) The turnover regime where the trajectory, close to the Gaussian
fixed point at $g=\lambda =0$, begins to depart from the canonical
one and turns over to the separatrix which connects the Gaussian 
with the non-Gaussian fixed point $(g_*,\lambda _*)$. By definition, the 
coordinates of the turning point $T$ are $g_T$ and $\lambda _T$, and it is passed 
at the scale $k=k_T$.\\
iii) The growing $\Lambda$ regime where $G(k)$ is approximately constant but 
$\Lambda(k)$ runs proportional to $k^4$. \\
iv) The fixed point regime where the trajectory approaches the non-Gaussian fixed
point in an oscillating manner. Directly at the fixed point one has $g(k)\equiv g_*$
and $\lambda(k)\equiv\lambda _*$, and therefore $G(k) \propto k^{-2}$ and
$\Lambda (k) \propto k^2$ for $k \rightarrow\infty$. The non-Gaussian fixed point
is responsible for the nonperturbative renormalizability of the theory.\\ \\
The behavior of the trajectory in the extreme infrared is not yet known since the
Einstein-Hilbert approximation breaks down when $\lambda(k)$ approaches the value
1/2. A more general truncation is needed to approximate the RG trajectory
in that region. For this reason the classical region i) does not necessarily
extend to $k=0$, and we speak about ``laboratory" scales for values of 
$k \equiv k_{\rm lab}$ in the region where $G$ and $\Lambda$ are constant. The
Planck mass is then defined as
\begin{equation}
 m_{\rm Pl}\equiv 1/\sqrt{G(k_{\rm lab})}.
\end{equation}
  
In the regimes i), ii) and iii) the trajectory is well approximated by linearizing
the RG flow about the Gaussian fixed point. In terms of the dimensionful 
parameters one finds that in its linear regime $G(k)=$const and \cite{h3}
\begin{equation} \label{run}
 \Lambda(k)=\Lambda _0 \left [ 1+(k/k_T)^4 \right ]
\end{equation}
where $\Lambda _0$ is a constant. The corresponding dimensionless $\lambda
=\Lambda /k^2$ runs according to
\begin{equation}\label{run'}
 \lambda (k)=\Lambda _0 \left [ \left (\frac{1}{k}\right )^2+
 \left ( \frac{k}{k_T^2}\right )^2 \right ]
\end{equation}
Note that this function is invariant under the ``duality transformation"
$k \mapsto k_T^2/k$:
\begin{equation}\label{du}
 \lambda(k)=\lambda(k_T^2/k).
\end{equation}

For further details and a discussion of the other types of trajectories we refer
to \cite{frank1} and \cite{h3}. The analysis in the following sections of this 
paper refers entirely to trajectories of Type IIIa.

The rest of this paper is organized as follows. After discussing the concept of
scale dependent distances in Section 2 we introduce, in Section 3, a specific
model of a fractal spacetime, the four-sphere in the sense of QEG. On this spacetime
all the constructions reviewed in the Introduction can be performed explicitly. 
In particular, in Section 4, we determine the corresponding cutoff modes, the 
set {\sf COM}($k$), and derive their proper coarse graining scale $\ell$. While $\ell$
can become arbitrarily small we shall demonstrate in Section 5 that the angular 
resolution of the COMs cannot be increased beyond a certain point, rendering the 
$S^4$ of QEG a kind of ``fuzzy sphere". In Section 6 we analyze this fuzzyness 
from the point of view of a macroscopic observer. Then, in Section 7, we discuss 
the idea of an ``intrinsic" distance of two spacetime points, as well as a new kind of
duality transformations exchanging large and small distances. Section 8 contains a brief
summary of the results.
\\

\noindent{\bf\large 2. Scale dependent distances}

With a fixed classical metric, the (geodesic, say) distance of two given points
$x$ and $y$ on a Riemannian manifold reads
\begin{equation}
 L(x,y)\equiv \int _{{\cal C}_{xy}}\left (g_{\mu\nu}dx^\mu dx^\nu \right )^{1/2}
\end{equation}
where ${\cal C}_{xy}$ denotes the geodesic
connecting $x$ to $y$. (For simplicity, we assume that $x$
and $y$ are close enough so that there are no caustics.)
In a quantum spacetime, $g_{\mu\nu}$ is replaced by the set of metrics
$\{\left < g_{\mu\nu}\right >_k \}$. As a result, the distance from $x$ to $y$
depends on $k$ now:
\begin{equation}\label{intlen}
 L_k(x,y)\equiv\int _{{\cal C}_{xy}}\left (\left < g_{\mu\nu}\right >_k
 dx^\mu dx^\nu \right )^{1/2}.
\end{equation}
The interpretation of this $k$-dependent distance is as follows. If $k$ parametrizes
the ``resolution of the microscope" with which the spacetime is observed, the
metric $\left < g_{\mu\nu}\right >_k$ and correspondingly the distance $L_k(x,y)$
pertain to a specific scale of resolution, and different observers, using 
microscopes of different $k$-values, will measure different lengths in general. 
This $k$-dependence of lengths is analogous to the ``coastline of England 
phenomenon" well known from fractal geometry \cite{mandel,avra}. What complicates
the situation in QEG is that $k$ has no direct physical meaning 
a priori and can be related
to an averaging scale only indirectly, for instance via the properties
of the COMs.\\

\noindent{\bf\large 3. The QEG four-sphere}

The QEG four-sphere is a manifold in the QEG sense, i.e. supplied with a 
family of infinitely many metrics $\{\left < g_{\mu\nu}\right >_k|k=0,\cdots,
\infty \}$. To be specific, it is the family of maximally symmetric solutions
of (\ref{fieldeq}) with positive curvature. It exists only provided $\Lambda(k)>0$,
which is the case for all type IIIa trajectories.
Strictly speaking, as long as we are restricted to the Einstein-Hilbert truncation, 
the family of metrics starts not at $k=0$, but at some $k=k_{\rm min}$, beyond which
(towards the infrared) the trajectory cannot be continued.

We may parametrize the four-sphere by coordinates $(\zeta,\eta,\theta,\phi)$
with ranges $0 < \zeta,\eta,\theta < \pi$ and $0 \leq \phi < 2 \pi$. The metric
can be written as
\begin{equation}\label{metr}
 \left < g_{\mu\nu}\right > _k dx^\mu dx^\nu =r^2(k)\left [ d \zeta ^2 +\sin ^2 \zeta 
 \left (
 d \eta ^2 + \sin ^2 \eta ( d \theta ^2 + \sin ^2 \theta d \phi ^2 )\right ) \right ],
\end{equation}   
where $r(k)$ is the $k$-dependent radius of the $S^4$. To determine it, one has to 
insert the ansatz (\ref{metr}) into the effective vacuum field equation (\ref{fieldeq}).
The result is 
\begin{equation}\label{rsol}
 r(k)=\sqrt{3/ \Lambda (k)}.
\end{equation}

The family of metrics (\ref{metr}), (\ref{rsol}) constitutes a concrete example of a
quantum spacetime as it was discussed in ref.~\cite{oliverfrac}. In particular it
has a scale dependent, in general non-integer fractal dimension. Both the spectral 
dimension ${\cal D}_S$ and the one based upon the anomalous dimension $\eta _N$
interpolate between 4 at macroscopic distances and 2 on microscopic scales.
Contrary to a Brownian curve or the coastline of England the fractal dimension 
of the quantum spacetime is equal or {\it smaller} than the classical one. The 
reason is that, according to (\ref{metr}) and (\ref{rsol}), distances {\it decrease}
when we {\it increase} the cutoff $k$. The metric scales as 
$\left < g_{\mu\nu}\right >_k \propto 1/\Lambda(k)$ so that in the fixed point 
regime $\left < g_{\mu\nu}\right >_k \propto 1/k^2$ implying $L_k(x,y)\propto 1/k$
for any (geodesic) distance.

In fact, on the sphere it is easy to write down the geodesic distance (\ref{intlen})
explicitly. Without loss of generality we may assume that the two points $x$ and $y$
are both located on the equator $\zeta =\eta =\theta =\pi /2$. Denoting their
$\phi$-angles by $\phi(x)$ and $\phi(y)$, respectively, and exploiting that on the
equator
\begin{equation}\label{equa}
 \left < g_{\phi\phi}\right >_k =r^2(k)=3/\Lambda(k)
\end{equation}
the geodesic distance $L_k(x,y)=\sqrt{\left < g_{\phi\phi}\right >}\int d \phi
=r(k)\int d \phi$ reads
\begin{eqnarray}\label{geodist}
 L_k(x,y)&=& \sqrt{3/\Lambda(k)}\; | \phi(x)-\phi(y)|\\ \nonumber
 &=& \frac{1}{k}\sqrt{\frac{3}{\lambda(k)}}\; | \phi(x)-\phi(y)|.
\end{eqnarray}
We shall come back to this expression later on.\\

\noindent {\bf\large 4. The cutoff modes and their resolving power}

On the quantum $S^4$, the scalar eigenfunctions of $-{\bf \Delta} _k$ have the discrete 
eigenvalues
\begin{equation}\label{evalues}
 {\cal E}_n=\frac{n(n+3)}{r^2(k)},\quad n=0,1,2,3,\cdots
\end{equation} 
with degeneracies
\begin{equation}
 D_n=\frac{1}{6}(n+1)(n+2)(2n+3).
\end{equation}
The eigenmodes are labeled by four integer quantum numbers $n$, $l_1$, $l_2$ and $m$,
where $n \geq l_1 \geq l_2 \geq |m|$. They have the form
\begin{equation}
 Y_{n l_1 l_2 m}(\zeta,\eta,\theta,\phi)= \; _4 \bar{P}^{l_1}_n (\zeta) \;
 _3 \bar{P}^{l_2}_{l_1}(\eta) \; _2 \bar{P}^{m}_{l_2}(\theta)\;
 \frac{1}{\sqrt{2 \pi}}e^{im \phi}.
\end{equation}
Here $_i \bar{P}^j_k(\alpha)$ are generalized associated Legendre functions, for details
see e.g. ref.~\cite{sphere}.

Let us determine the associated set of cutoff modes {\sf COM}($k$), i.e. 
the eigenfunctions with $-{\bf \Delta} _k$-eigenvalue 
as close as possible to $k^2$. Inserting ${\cal E}\approx k^2$ into 
(\ref{evalues}) and using eq.~(\ref{rsol}) for $r(k)$, we find the following
equation for the $n$-quantum number of the COMs at scale $k$:
\begin{equation}\label{kev}
 n(k)\left ( n(k)+3 \right )\approx\frac{3k^2}{\Lambda(k)}=\frac{3}{\lambda(k)}.
\end{equation} 
The eigenvalues for the vector and
tensor modes are slightly different, but for large $n$ the 
difference becomes negligible and the spectrum is to a good approximation continuous.
We will use this continuum approximation 
since we are interested in small angular distances $\Delta\phi$
anyway. Then eq.(\ref{kev}) becomes
\begin{equation}\label{nofk}
 n(k) \approx\sqrt{\frac{3}{\lambda(k)}}.
\end{equation} 
Obviously $n(k)$ is indeed large if $\lambda(k)\ll 1$.

Since the eigenvalue belonging to $Y_{nl_1l_2m}$ depends on $n$ only, the set 
{\sf COM}($k$)
consists of all harmonics $Y_{nl_1l_2m}$ with $n$ fixed by eq.~(\ref{nofk}) and the
other quantum numbers $l_1$, $l_2$ and $m$ arbitrary except for the constraint
$n \geq l_1 \geq l_2 \geq |m|$. If $\lambda(k)\ll 1$, the dimension of {\sf COM}($k$)
as a vector space is 
approximately $n(k)^3 /3 \approx \sqrt{3}\lambda(k)^{-3/2}\gg 1$.

Apart from its obvious dependence on the scale, the set {\sf COM}($k$) depends on the RG
trajectory via the function $\lambda(k)$ which determines $n(k)$. It is 
important to note that the function $\lambda =\lambda(k)$ is not invertible
in general and that different $k$'s can lead to the same {\sf COM}($k$).
Let us look at the Type IIIa trajectory in Fig.~1 as an example. First we concentrate
on its part close to the turning point, staying away from the spiraling regime
in the UV, and the IR region where the Einstein-Hilbert truncation breaks down.
We observe then that for every scale $k<k_T$ below the turning point there exists
a corresponding scale $k^\sharp >k_T$ which has the same $\lambda$- and therefore
$n$-value. As a result, the corresponding cutoff modes are equal at the two scales:
{\sf COM}($k$)={\sf COM}($k^\sharp$).

If the turning point is sufficiently close to the Gaussian fixed point, 
and $k$ is not too far from $k_T$, we may use the linearization (\ref{run'})
for an approximate determination of $k^\sharp$. Because of the ``duality symmetry"
(\ref{du}) it is given by
\begin{equation}\label{ksharp}
 k^\sharp =k_T^2/k.
\end{equation}  
Thus, in the linear regime, {\sf COM}($k$)={\sf COM}($k_T^2/k$).

Outside the linear regime the relation between $k^\sharp$ and $k$ is more implicated.
The situation becomes even more involved in the regime close to the non-Gaussian
fixed point. There, because of the spiral structure, very many different $k$-values
have the same $\lambda(k)$ and {\sf COM}($k$).
 
Next we analyze the degree of position dependence of the functions in {\sf COM}($k$)
and try to quantify their ``resolving power". In order to convert the 
estimate for $n(k)$, eq.~(\ref{nofk}), to
an angular resolution we note that it is sufficient
to do so for one position and one direction. By the translation and rotation symmetries
of the sphere, the resolution will be the same at any other point and in any other
direction. We therefore choose to determine the angular resolution of the modes
along the equator $\zeta =\eta =\theta =\pi /2$.

Two of the ${\bf \Delta} _k$-eigenfunctions with eigenvalue $n(k)$,
namely $Y_{\pm}\equiv Y_{nnn \pm n}$, oscillate most rapidly as a function of $\phi$,
and we shall use them in order to define the angular resolution. 
Their $\phi$-dependence is $e^{\pm in \phi}$ and
the corresponding angular resolution is
\begin{equation}\label{dphi}
 \Delta\phi(k)=\frac{\pi}{n(k)}=\pi\sqrt{\frac{\lambda(k)}{3}}.
\end{equation}
As expected, the angular resolution implied by the COMs depends on the RG trajectory.
It does so only via the function $\lambda = \lambda(k)$ and, as a result,
can be of the same size for different values of $k$. In particular
$\Delta\phi(k)=\Delta\phi(k^\sharp)$ and, in the linear regime, 
$\Delta\phi(k)=\Delta\phi(k_T/k^2)$.

The angular separation (\ref{dphi}) is the {\it coordinate} distance of two 
consecutive zeros of the real or imaginary part of $Y_\pm$ along the equator 
$\zeta =\eta =\theta =\pi /2$. By definition, the
COM scale $\ell$ is the {\it proper} length corresponding to $\Delta\phi(k)$ 
as computed with the metric $\left < g_{\mu\nu}\right >_k$ of eqs.~(\ref{metr}),
(\ref{rsol}). It is given by $\ell(k)=\Delta\phi(k)\sqrt{\left < g_{\phi\phi}
\right >_k}$ or, using eq.~(\ref{equa}), $\ell(k)=\Delta\phi(k)\sqrt{3/\Lambda(k)}
=(1/k)\Delta\phi(k)\sqrt{3/\lambda(k)}$. Hence, with (\ref{dphi}),
\begin{equation}\label{klength'}
 \ell(k)=\frac{\pi}{k}.
\end{equation}  
So we find that, as in theories on a classical flat spacetime, the natural 
proper length scale $\ell$ of the {\sf COM}($k$)-modes is just $\pi/k$.
Thanks to the symmetry of the sphere it is neither position nor
direction dependent. \\

\noindent{\bf\large 5. Lower bound on the angular resolution}

Taking just the 
(perhaps expected) result $\ell\propto 1/k$, 
it seems as if nothing remarkable had happened.
But the surprising effects appear in our result for the angular resolution, 
eq.~(\ref{dphi}). As we can see from the flow diagram of Fig.~1, $\lambda(k)$ takes
on a minimum value $\lambda _T$ at the turning point $T$.
Only the Type IIIa trajectories have this turning point, and this is 
one of the features that makes
them particularly interesting. In fact, as $\lambda(k)\geq\lambda _T$ for any scale
$k$, we conclude that the angular resolution $\Delta\phi (k)$ is bounded below
by the minimum angular resolution
\begin{equation}\label{dephmin}
 \Delta\phi _{\rm min}=\pi\sqrt{\frac{\lambda _T}{3}}.
\end{equation}
Stated differently, there 
does not exist any cutoff $k$ for which $\Delta\phi(k)$ would be smaller
than $\Delta\phi _{\rm min}$. On the other hand, angular resolutions between 
$\Delta\phi _{\rm min}$ and $\Delta\phi _*$, defined by
\begin{equation}
 \Delta\phi _*\equiv\pi\sqrt{\frac{\lambda _*}{3}},
\end{equation}
are realized for at least two scales $k$, one of them above, the other below $k_T$.

What has happened here? Coming from small $k$, we travel along the RG trajectory 
and follow its spherical solutions, 
observing spacetime with a ``microscope" of variable proper resolution $\ell(k)$.
At first, in the classical regime, an increase of $k$ leads to the resolution of
finer and finer structures since $\Lambda =$const implies $\Delta\phi(k)
\propto 1/k$. 
For the canonical RG trajectory, this behavior would
continue even for $k \rightarrow\infty$. 
In quantum gravity, however, in region ii), the sphere starts to {\it shrink},
due to a growing cosmological constant $\Lambda (k)$. At the turning point scale $k_T$ 
at which $\lambda (k)$ assumes its minimum
$\lambda _T$, the shrinking becomes faster than the improvement of the resolution
($r(k)\propto k^{-2}$ in region iii)). 
Although we can resolve smaller and smaller proper distances, this is of
no use, since the sphere is shrinking so fast that a {\it smaller} proper length 
corresponds to a {\it larger} angular distance. Finally, in the fixed point regime
(at large angles although this is an ultraviolet fixed point!),
the shrinking slows down to a rate that cancels exactly the improved resolution
of the microscope ($r(k)\propto k^{-1}$) so that the angular resolution
approaches a constant value $\Delta\phi _*$ after the oscillations have been damped 
away. (Depending on the value of $\lambda _*$ it may even happen that the UV
fixed point corresponds to so large angular scales that the 
approximation $n \gg 1$ breaks down.)

The minimum of $\Delta\phi$ at the turning point is equivalent to a maximum of the $n$
quantum number the COMs can have:
\begin{equation}
 n_{\rm max}\approx\sqrt{\frac{3}{\lambda _T}}.
\end{equation}
Does this mean that in the fundamental path integral underlying the flow equation
not all quantum fluctuations are integrated out when $k$ is lowered from infinity
to $k=0$? Are the modes with $n>n_{\rm max}$ left out? The answer to these questions 
is a clear no. One has to carefully distinguish the quantum metric $\gamma _{\mu\nu}$,
the variable of integration in the path integral $\int {\cal D}\gamma 
\exp \left ( -S[\gamma]+...\right )$, from its expectation value, or mean field,
$\left < g_{\mu\nu}\right >_k$. As we explained above, the 
shifted functional integration
over $h_{\mu\nu}=\gamma _{\mu\nu}-\bar{g}_{\mu\nu}$ which is implicit in the definition of
$\Gamma _k \left [ g_{\mu\nu},\bar{g}_{\mu\nu}\right ]$ is organized according to the
eigenmodes of the covariant Laplacian $\bar{D}^2$ constructed from the arbitrary
but {\it fixed} background metric $\bar{g}_{\mu\nu}$. This integration is performed 
before $g_{\mu\nu}$ and $\bar{g}_{\mu\nu}$ are identified and the result is equated
to $\left < g_{\mu\nu}\right >_k$. At scale $k$, the $-\bar{D}^2$ modes down
to eigenvalues $\approx k^2$ are integrated out, and therefore on the way from 
$k=\infty$ to $k=0$ it is really {\it all} modes of 
$h_{\mu\nu}$ and therefore $\gamma _{\mu\nu}$ that are 
integrated out since the eigenfunctions of $\bar{D}^2$ form a complete
system.

The results of the present paper should instead be thought of as reflecting properties
of the mean field $\left < g_{\mu\nu}\right >_k$. Rather than the spectrum of the
$k$-independent operator $\bar{D}^2$ we analyzed that of the 
explicitly $k$-dependent Laplacian $D^2 \left ( \left < g_{\mu\nu}\right >_k \right )$;
its explicit $k$-dependence is due to the scale dependence of the on-shell metric,
the solution to the effective Einstein equation. Our argument reveals that the 
effective spacetime with the running on-shell metric cannot support harmonic modes
of arbitrarily fine angular resolution.

This phenomenon is a purely dynamical one; the finite resolution is not built in 
at the kinematical (i.e. $\gamma _{\mu\nu}$-) level, as it would be the case,
for instance, if the fundamental theory was defined on a lattice. It is also 
important to stress that, if the non-Gaussian fixed point exists, the 
Green's functions $G_n(x_1,x_2,\cdots,x_n)$
can be made well defined at all non-coincident points, i.e. 
for arbitrarily small coordinate distances among the $x_i^\mu$'s. Those
Green's functions contain information even about angular scales smaller than 
$\Delta\phi _{\rm min}$, in particular they ``know" about the asymptotic safety 
of the theory which manifests itself only at scales $k \gg k_T$.

In fact, the argument leading to the finite resolution $\Delta\phi _{\rm min}$
is fairly independent of the high energy behavior of the theory. The crucial 
ingredient in the above reasoning was the occurrence of a minimum value 
for $\lambda(k)$. This minimum occurs as a direct consequence of the $k^4$-running
of $\Lambda (k)$ given in eq.~(\ref{run}); how the trajectory continues 
beyond the scaling regime of the Gaussian fixed point was not important.
However, this $k^4$-running occurs already in standard perturbation theory, simply 
reflecting the quartic divergences of all vacuum diagrams. From this point of
view our argument is rather robust; it would apply even to a non-asymptotically 
safe theory in which the impact of the perturbative $k^4$-running is taken seriously.

Moreover, while the above argument was formulated within the Einstein-Hilbert 
truncation, it is likely to be still valid in more general truncations and
the exact theory. The only prerequisite is that, near $k=k_T$, the effective
field equations must admit $S^4$ solutions. For some of the trajectories this will
always be the case presumably.

The upper bound on the angular momentum like quantum number $n$ is reminiscent
of the ``fuzzy sphere" constructed in ref.~\cite{fuzzy}. While in the case of 
the fuzzy sphere the finite angular resolution is put in ``by hand", in the present case 
it emerges as a consequence of the quantum gravitational dynamics .\\

\noindent{\bf\large 6. Minimum proper length for a macroscopic observer}

A macroscopic, classical observer would find it natural to employ the metric
\begin{equation}
 g_{\mu\nu}^{\rm macro}=\left < g_{\mu\nu}\right >_{k_{\rm lab}},
\end{equation}   
where $k_{\rm lab}$ is any scale in the classical regime in which $G$ and $\Lambda$
do not run (in Fig.~1 between the 
points $P_1$ and $P_2$). Using this metric we can define a
``macroscopic distance" for any two given points $x$ and $y$:
\begin{equation}
 L^{\rm macro}(x,y)\equiv \int _{{\cal C}_{xy}}
 \left (g_{\mu\nu}^{\rm macro} dx^\mu dx^\nu \right )^{1/2}.
\end{equation} 

It is instructive to ask which proper length the macroscopic observer using
$g_{\mu\nu}^{\rm macro}$ would ascribe to the finite coordinate distance
$\Delta\phi _{\rm min}$. Denoting this proper length by $L_{\rm min}^{\rm macro}$
we obtain
\begin{equation}
 L_{\rm min}^{\rm macro}=r(k_{\rm lab})\Delta\phi _{\rm min}
 =\left [ 3/\Lambda (k_{\rm lab})\right ] ^{1/2}\Delta\phi _{\rm min},
\end{equation}
and with eq.~(\ref{dephmin}),
\begin{equation}\label{lmami}
 L_{\rm min}^{\rm macro}
 =\frac{\pi}{k_T}\sqrt{\frac{\Lambda(k_T)}{\Lambda(k_{\rm lab})}}
 =\frac{\pi}{k_T}\left [\frac{2}{1+(k_{\rm lab}/k_T)^4}\right ]^{1/2}.
\end{equation}
In the second equality of (\ref{lmami}) we assumed that the turning point is
sufficiently close to the Gaussian fixed point so that eq.~(\ref{run}) is a
good approximation for $\Lambda(k)$. (For the trajectory which seems realized 
in Nature this is indeed the case.) If $k_{\rm lab}\ll k_T$ we find that
$L_{\rm min}^{\rm macro}$ is essentially the same as the turning point scale:
\begin{equation}\label{lmami2}
 L_{\rm min}^{\rm macro}=\sqrt{2}\pi k_T^{-1}.
\end{equation}

Remarkably, this minimal proper length is different in general from the Planck length
which is usually thought to set the minimal length scale. Using the linearized 
flow one can show \cite{h3} that $k_T/m_{\rm Pl}\approx\sqrt{g_T}$. So, if $g_T$ is 
small, $L_{\rm min}^{\rm macro}$ can be much larger than $\ell _{\rm Pl}
\equiv m_{\rm Pl}^{-1}$. The trajectory realized in Nature seems to be an extreme
example: It has $g_T \approx 10^{-60}$, implying $k_T^{-1}\approx 10^{30}\ell _{\rm Pl}
\approx 10^{-3}$ cm, and $L_{\rm min}^{\rm macro}$ is of the same order of
magnitude.

Should we therefore expect to find an $L_{\rm min}^{\rm macro}$ 
of the order of $10^{-3}$ cm in the real world?
The answer is no, most probably. The reason is that our present discussion 
is based upon the vacuum field equations where it is the value of the cosmological
constant alone which determines the curvature of spacetime. In presence of matter,
the scale dependence of $\Lambda$ can have an observable effect only if the 
vacuum energy density $\rho _\Lambda \equiv \Lambda /8 \pi G$ is comparable
to the matter energy density (including the matter energy density of the measuring
device). \\

\noindent{\bf \large 7. Intrinsic distance and scale doubling}

In fractal geometry and any framework involving a length scale dependent metric
one can try to define an ``intrinsic" distance of any two points $x$ and $y$ 
by adjusting the resolving power of the
``microscope" in such a way that the length scale it resolves equals 
approximately the, yet to be determined, intrinsic (geodesic) distance from $x$ to $y$.
\footnote{This kind of dynamical adjustment of the resolution has also been used in the 
RG improvement of black hole \cite{bh} and cosmological \cite{cosmo1}-\cite{mof}
spacetimes, see in particular ref.~\cite{h2}.}

To be concrete, let us fix two points $x$ and $y$ and let us try to assign to them
a cutoff scale $k \equiv k(x,y)$ which satisfies
\begin{equation}\label{self}
 L_{k(x,y)}(x,y)=\ell(k(x,y)).
\end{equation}
Eq.~(\ref{self}) is a self-consistency condition for $k(x,y)$: the LHS of (\ref{self})
is the distance from $x$ to $y$ as seen by a microscope with $k=k(x,y)$, and the
RHS is precisely the resolution of this microscope. If (\ref{self}) has a unique
solution $k(x,y)$ one defines the intrinsic distance by setting
\begin{equation}\label{lin}
 L_{\rm in}(x,y)\equiv L_{k(x,y)}(x,y).
\end{equation}
Since $\ell(k)=\pi/k$, this distance is essentially the inverse cutoff scale:
\begin{equation}\label{link}
 L_{\rm in}(x,y)=\frac{\pi}{k(x,y)}.
\end{equation}

Let us evaluate the self-consistency condition (\ref{self}). Without loss of 
generality we may assume again that $x$ and $y$ are located on the equator of $S^4$
so that eq.~(\ref{geodist}) applies. Then, by virtue of (\ref{klength'}), 
eq.~(\ref{self}) boils down to the following implicit equation for $k(x,y)$:
\begin{equation}\label{lamk}
 \lambda(k(x,y))=\frac{3}{\pi ^2}| \phi(x)-\phi(y)|^2.
\end{equation}    

Recalling the properties of the function $\lambda(k)$ for a Type IIIa trajectory
we see that (\ref{lamk}) does {\it not} admit a unique solution for $k(x,y)$.
If $x$ and $y$ are such that $| \phi(x)-\phi(y)|<\Delta\phi _{\rm min}$ it 
possesses no solution at all, and if $| \phi(x)-\phi(y)|>\Delta\phi _{\rm min}$
it has at least two solutions. Staying away from the deep UV and IR regimes,
every solution $k(x,y)<k_T$ on the lower branch of the RG trajectory has 
a partner solution $k(x,y)^\sharp >k_T$ on its upper branch. As a result, the 
intrinsic distance of $x$ and $y$ is either undefined, or there exist at least two 
different lengths which satisfy the self-consistency condition (\ref{self}).

In the linear regime where $k^\sharp =k_T^2/k$, the two lengths $L_{\rm in}(x,y)
=\pi /k(x,y)$ and $L_{\rm in}(x,y)^\sharp =\pi / k(x,y)^\sharp$ are related by
\begin{equation}\label{ldual}
 L_{\rm in}(x,y)^\sharp =\frac{L_T^2}{L_{\rm in}(x,y)}
\end{equation}
where $L_T \equiv \pi / k_T$. Eq.~(\ref{ldual}) is reminiscent of the dualities
in string theory. If $L_{\rm in}(x,y)$ is large compared to the turning point length
scale $L_T$, the ``dual" scale $L_{\rm in}(x,y)^\sharp$ is small. In the extreme
case, when applied to Nature's RG trajectory, the duality (\ref{ldual}) would
even exchange the Planck- with the Hubble-regime: $L_{\rm in}(x,y)\approx H_0^{-1}$
implies $L_{\rm in}(x,y)^\sharp \approx l_{\rm Pl}$.

Thus we found one more instance where a quite unexpected ``doubling" of $k$-scales
makes its appearance. The first phenomenon of this kind which we encountered 
was the exact equality of {\sf COM}($k$) with {\sf COM}($k^\sharp$),
and later on we saw that also the angular resolution $\Delta\phi$ is exactly the
same at $k$ and $k^\sharp$. We shall refer to this phenomenon as a ``scale doubling",
keeping in mind however that in the spiraling regime there are many more $k$-values
with the same {\sf COM}($k$) and $\Delta\phi(k)$.

At a formal level the origin of the doubling is easy to understand. It is due to
the ``back bending" of the RG trajectory at the turning point $T$ which implies 
that the function $\lambda =\lambda(k)$ assumes a minimum at a finite scale $k=k_T$.
Only the trajectories of Type IIIa possess a turning point of this kind, and this
is one of the reasons 
why they are particularly interesting and we restricted our discussion to them.
It is important to understand that the occurrence of this turning point is a pure
quantum effect. Switching off the renormalization (i.e. quantum) effects, 
the Type IIIa trajectory is substituted by the canonical trajectory shown in Fig.~1
which has no turning point and no scale doubling therefore.

While its origin is quite clear, the physical implications of the scale doubling
and the duality symmetry are somewhat mysterious. To some extent the difficulty
of giving a precise physical meaning to them is related to the fact that one actually
should define the ``resolution of the microscope" in terms of realistic experiments
rather than the perhaps too 
strongly idealized mathematical model of a measurement based upon the
COMs. For various reasons it seems premature to assign a direct observational 
relevance to the minimal angular resolution and the scale doubling:\\ \\
(i) We found that only the {\it coordinate} distance $\Delta\phi(k)$ assumes a
minimum, but not the corresponding {\it proper} distance computed with the 
running metric $\left < g_{\mu\nu} \right >_k$. In particular the resolution
function $\ell(k)=\pi /k$ is exactly the same as in flat space. (But nevertheless
the COM-microscope is unable to distinguish points with an angular separation
below $\Delta\phi _{\rm min}$ !)\\
(ii) Our analysis applies to pure gravity. In presence of matter the ``fuzziness"
of the $S^4$ can become visible probably only at scales where the cosmological 
constant dominates the energy density.\\
(iii) As a special case of (ii), the fuzziness might be masked by the backreaction
of a realistic (i.e. gravitating) measuring apparatus on the spacetime structure.\\
(iv) As for a possible physical significance of the duality symmetry it is to be
noted that the two scales which it relates, $k<k_T$ and $k^\sharp >k_T$,
have a rather different status as far as quantum fluctuations about the mean field 
metric $\left < g_{\mu\nu}\right >_k$ are concerned. The structure of the exact
RG equation is such that the fluctuations are the larger the stronger the 
renormalization effects are. As a result, the metric fluctuations about 
$\left < g_{\mu\nu}\right >_{k^\sharp}$ on the upper branch are certainly larger
than at the dual point on the lower branch of the RG trajectory.\\ \\
Clearly more work is needed in order to understand these issues better. We
emphasize, however, that the above discussion deals only with the {\it interpretation}
of QEG, and not with its basic {\it construction} along the lines of the asymptotic
safety scenario. As we pointed out already, the existence of a finite 
$\Delta\phi _{\rm min}$ is perfectly consistent with having integrated out all
modes of the quantum metric. Throughout the present paper we assumed that the quantum
theory has been constructed already and that its RG trajectories are known.
What we did is to derive the running mean field metric for a given trajectory and 
to analyze its properties, in particular with respect to the running set of 
cutoff modes {\sf COM}($k$) it gives rise to. 
\\

\noindent{\bf\large 8. Summary}

In this paper we constructed a special example of a ``QEG spacetime", the quantum
4-sphere, and analyzed its properties. We were particularly interested in the
relationship between the IR cutoff $k$ and the coarse graining scale $\ell$.
Strictly speaking this scale, the resolving power of the ``microscope" with which 
we observe the spacetime structure, depends in a complicated way on how this 
``microscope" is realized in practice, i.e. on what kind of experiment we perform
in order to probe spacetime. Rather than analyzing possible experiments explicitly
we used a simple and natural mathematical model for the resolving power, namely
we defined $\ell$ to be the typical proper length scale of the cutoff modes.
The main motivations for this definition are that on a classical spacetime it
reproduces the standard result, and that it is an intrinsic property of any
RG trajectory: given the trajectory $k \mapsto \Gamma _k$, we can solve for
$\left < g_{\mu\nu}\right >_k$ at each scale, build ${\bf \Delta} _k$ from it,
and solve its eigenvalue problem to find {\sf COM}($k$).

Applying this algorithm to the quantum $S^4$ we found that the IR cutoff and the 
proper coarse graining scale are related in exactly the same way as in classical
flat space: $\ell(k)=\pi /k$. In obtaining this result it was crucial to employ
the {\it running} metric $\left < g_{\mu\nu}\right >_k$ for converting 
coordinate to proper distances.\\ 
\indent Regarding the question as to whether QEG generates a minimal length dynamically,  
again, one should in principle analyze realistic experiments on a case-by-case 
basis. As a first step in this direction we analyzed this issue within the 
COM-model of the ``microscope". The results are:\\ \\
(a) There is no lower bound on proper distances measured with the running metric
$\left < g_{\mu\nu}\right >_k$.\\
(b) There is a nonzero minimal angular separation $\Delta\phi _{\rm min}$, i.e.
a minimal coordinate distance that the COMs can resolve.\\
(c) There is a nonzero lower bound $L_{\rm min}^{\rm macro}$ on proper distances
measured with a fixed  macroscopic metric.\\ \\
The statements (a) and (b) above can be true simultaneously only thanks to the running
of the gravitational parameters: Increasing $k$ the on-shell spacetime shrinks,
and this effect counteracts our ability to separate two points by making $k$
larger. This is the physical origin of the finite angular resolution 
$\Delta\phi _{\rm min}$. Converting $\Delta\phi _{\rm min}$ to a proper length
with the running metric $\left < g_{\mu\nu}\right >_k$, the resulting
length becomes arbitrarily small for $k \rightarrow\infty$. Using a fixed
(macroscopic) metric instead, the finite angular resolution amounts to a finite
minimal proper distance $L_{\rm min}^{\rm macro}$, however. 
\\ \\
{\bf Acknowledgement:} We would like to thank A.~Bonanno, O.~Lauscher and R.~Percacci 
for helpful discussions.


\newpage

\begin{thebibliography}{minangle}
\bibitem{garay} 
For a review see
L.J.~Garay, Int.\ J.\ Mod.\ Phys.\ A10 (1995) 145 and references therein.
%
\bibitem{wein}
S.~Weinberg
in \textit{General Relativity, an Einstein Centenary Survey},
S.W.~Hawking and W.~Israel (Eds.),
Cambridge University Press (1979);
S.~Weinberg,
hep-th/9702027.

\bibitem{mr}
M.~Reuter,
Phys.\ Rev.\ D 57 (1998) 971 and hep-th/9605030.
%
\bibitem{percadou}
D.~Dou and R.~Percacci,
Class.\ Quant.\ Grav.\ 15 (1998) 3449.
%
\bibitem{oliver1}
O.~Lauscher and M.~Reuter,
Phys.\ Rev.\ D 65 (2002) 025013 and\\ hep-th/0108040.
%
\bibitem{frank1}
M.~Reuter and F.~Saueressig,
Phys.\ Rev.\ D 65 (2002) 065016 and\\ hep-th/0110054.
%
\bibitem{oliver2}
O.~Lauscher and M.~Reuter, Phys.\ Rev.\ D 66 (2002) 025026 and\\
 hep-th/0205062.
%
\bibitem{oliver3}
O.~Lauscher and M.~Reuter,
Class.\ Quant.\ Grav.\ 19 (2002) 483 and\\ hep-th/0110021.
%
\bibitem{oliver4}
O.~Lauscher and M.~Reuter,
Int.\ J.\ Mod.\ Phys.\ A 17 (2002) 993 and\\ hep-th/0112089.
%
\bibitem{souma}
W.~Souma,
Prog.\ Theor.\ Phys.\ 102 (1999) 181.
%
\bibitem{perper1}
R.~Percacci and D.~Perini,
Phys.\ Rev.\ D 67 (2003) 081503.
%
\bibitem{perper2}
R.~Percacci and D.~Perini,
Phys.\ Rev.\ D 68 (2003) 044018.
%
\bibitem{perini}
D.~Perini, Nucl.\ Phys.\ Proc.\ Suppl.\ 127 C (2004) 185.
%
\bibitem{frank2}
M.~Reuter and F.~Saueressig,
Phys.\ Rev.\ D 66 (2002) 125001 and\\ hep-th/0206145;
Fortschr.\ Phys.\ 52 (2004) 650 and hep-th/0311056.
%
\bibitem{litimgrav}
D.~Litim, Phys.\ Rev.\ Lett.\ 92 (2004) 201301.
%
\bibitem{prop}
A.~Bonanno, M.~Reuter, JHEP\ 02 (2005) 035 and hep-th/0410191.
%
\bibitem{essential}
R.~Percacci and D.~Perini, hep-th/0401071.
%
\bibitem{hier}
R.~Percacci, hep-th/0409199.
%
\bibitem{max}
P.~Forg\'acs and M.~Niedermaier,
hep-th/0207028; \\
M.~Niedermaier,
JHEP 12 (2002) 066;
Nucl.\ Phys.\ B 673 (2003) 131.
%
\bibitem{avact}
C.~Wetterich,
Phys.\ Lett.\ B 301 (1993) 90.
%
\bibitem{ym}
M.~Reuter and C.~Wetterich,
Nucl.\ Phys.\ B 417 (1994) 181,
Nucl.\ Phys.\ B 427 (1994) 291,
Nucl.\ Phys.\ B 391 (1993) 147,
Nucl.\ Phys.\ B 408 (1993) 91; 
M.~Reuter,
Phys.\ Rev. D 53 (1996) 4430,
Mod.\ Phys.\ Lett.\ A 12 (1997) 2777.
%
\bibitem{avactrev}
For a review see:
J.~Berges, N.~Tetradis and C.~Wetterich,
Phys.\ Rep.\ 363 (2002) 223;
C.~Wetterich,
Int.\ J.\ Mod.\ Phys.\ A 16 (2001) 1951.
%
\bibitem{oliverfrac}
O.~Lauscher and M.~Reuter, JHEP\ 10 (2005) 050 and hep-th/0508202.

\bibitem{avra}
D.~ben-Avraham and S. Havlin, {\it Diffusion and reactions in fractals and 
disordered systems}, Cambridge University Press, Cambridge (2004).
%
\bibitem{ajl1}
J.~Ambj\o rn, J.~Jurkiewicz and R.~Loll,
Phys.\ Rev.\ Lett.\ 93 (2004) 131301.
%
\bibitem{ajl2}
J.~Ambj\o rn, J.~Jurkiewicz and R.~Loll,
Phys.\ Lett.\ B 607 (2005) 205.
%
\bibitem{ajl34}
J.~Ambj\o rn, J.~Jurkiewicz and R.~Loll,
preprints hep-th/0505113;\\ hep-th/0505154; hep-th/0509010.
%
\bibitem{nino} H.~Kawai, M.~Ninomiya, Nucl.\ Phys.\ B 336 (1990) 115;\\
R.~Floreanini and R.~Percacci, Nucl.\ Phys.\ B 436 (1995) 141;\\
I.~Antoniadis, P.O.~Mazur and E.~Mottola, Phys.\ Lett.\ B 444 (1998) 284.

\bibitem{back}L.F.~Abbott, Nucl.\ Phys.\ B 185 (1981) 189;
B.S.~DeWitt, Phys.\ Rev.\ 162 (1967) 1195;
M.T.~Grisaru, P.~van Nieuwenhuizen and C.C.~Wu, Phys.\ Rev.\ D 12 (1975) 
3203;
D.M.~Capper, J.J.~Dulwich and M.~Ramon~Medrano, Nucl.\ Phys.\ B 254 (1985)
737;
S.L.~Adler, Rev.\ Mod.\ Phys.\ 54 (1982) 729.
%
\bibitem{bagber}
For a review see:
C.~Bagnuls and C.~Bervillier,
Phys.\ Rep.\ 348 (2001) 91; 
T.R.~Morris,
Prog.\ Theor.\ Phys.\ Suppl.\ 131 (1998) 395;
J.~Polonyi, Central Eur.\ J.\ Phys. 1 (2004) 1.
%
\bibitem{h3}
M.~Reuter and H.~Weyer,
JCAP\ 12 (2004) 001
and hep-th/0410119.
%
\bibitem{mandel}
B.~Mandelbrot, {\it The Fractal Geometry of Nature,} Freeman,\\ New York (1983).

\bibitem{sphere}
M.~Rubin and C.~Ordonez, J.\ Math.\ Phys.\ 26 (1985) 65;\\
A.~Higuchi, J.\ Math.\ Phys.\ 28 (1987) 1553.

\bibitem{fuzzy} 
J.~Madore, Class.\ Quant.\ Grav. 9 (1992) 69.

\bibitem{bh}
A.~Bonanno and M.~Reuter,
Phys.\ Rev.\ D 62 (2000) 043008 and\\ hep-th/0002196;
Phys.\ Rev.\ D 60 (1999) 084011 and gr-qc/9811026.
%
\bibitem{cosmo1}
A.~Bonanno and M.~Reuter,
Phys.\ Rev.\ D 65 (2002) 043508 and\\ hep-th/0106133; M.~Reuter and 
F.~Saueressig, JCAP\ 09 (2005) 012 and hep-th/0507167.
%
\bibitem{cosmo2}
A.~Bonanno and M.~Reuter,
Phys.\ Lett.\ B 527 (2002) 9 and\\ astro-ph/0106468; 
Int.\ J.\ Mod.\ Phys.\ D 13 (2004) 107 and\\ astro-ph/0210472.
%
\bibitem{elo}
E.~Bentivegna, A.~Bonanno and M.~Reuter,
JCAP 01 (2004) 001 \\ and
astro-ph/0303150.
%
\bibitem{esposito}
A.~Bonanno, G.~Esposito and C.~Rubano,
Gen.\ Rel.\ Grav.\ 35 (2003) 1899;
Class.\ Quant.\ Grav.\ 21 (2004) 5005;
A.~Bonanno, G.~Esposito, C.~Rubano and P.~Scudellaro, astro-ph/0507670.
%
\bibitem{h1}
M.~Reuter and H.~Weyer,
Phys.\ Rev.\ D 69 (2004) 104022
and\\ hep-th/0311196.
%
\bibitem{h2}
M.~Reuter and H.~Weyer,
Phys.\ Rev.\ D 70 (2004) 124028
and \\ hep-th/0410117.
%
\bibitem{mof}
J.~Moffat,
JCAP\ 05 (2005) 003
and astro-ph/0412195; J.R.~Brownstein and J.~Moffat, astro-ph/0506370 and
astro-ph/0507222.
%
\end{thebibliography}
\end{document}